\documentclass[notoc,cits]{PoS}
\usepackage{xspace}
\usepackage{graphicx}
\usepackage{amssymb}
\usepackage{amsmath}
\usepackage{url}
\usepackage{rotating}
\usepackage{microtype}
\usepackage{natbib}
\usepackage[T1]{fontenc}
\usepackage{mathptmx}

\bibpunct{[}{]}{,}{a}{}{,}

\newcommand{\integral}{\textsl{INTEGRAL}\xspace}

\newcommand{\ca}{\mbox{$\sim$}}

\newcommand{\Rsun}{\ensuremath{R_\odot}\xspace}
\newcommand{\Msun}{\ensuremath{M_\odot}\xspace}

\newcommand{\vela}{\mbox{Vela~X-1}\xspace}

\newcommand{\hd}{HD\,77581\xspace}
\newcommand{\cps}{\ensuremath{\text{counts}\,\text{s}^{-1}}\xspace}

\title{Highly structured wind in Vela X-1}

\author{\mbox{\speaker{Ingo Kreykenbohm}$^{1,2}$}, \mbox{J\"orn
    Wilms$^{1,2}$}, \mbox{Peter Kretschmar$^{3}$}, \mbox{Jos\'e Miguel
    Torrej{\'o}n$^{4,5}$}, \mbox{Katja Pottschmidt$^{6,7}$},
  \mbox{Manfred Hanke$^{1,2}$}, \mbox{Andrea Santangelo$^{8}$},
  \mbox{Carlo Ferrigno$^{8,9,10}$}, and
  \mbox{R\"udiger Staubert$^{8}$}\\
  \llap{$^1$}Dr.  Karl Remeis-Sternwarte Bamberg, Sternwartstrasse~7,
  96049~Bamberg, Germany,\\\llap{$^2$} Erlangen Centre for
  Astroparticle Physics, Erwin-Rommel-Str. 1, 91058~Erlangen,
  Germany\\$^3$ European Space
  Astronomy Centre, Villafranca del Castillo, P.O. Box 78, 28691
  Villanueva de la Ca{\~ n}ada, Madrid, Spain\\ $^4$ Departamento de
  F\'isica, Ingenier\'ia de Sistemas y Teor\'ia de la Se\~nal, Escuela
  Polit\'enica Superior, Universidad de Alicante, Ap.\ 99, 03080
  Alicante, Spain\\$^5$ Kavli Institute for Astrophysics and Space
  Research, Massachusetts Institute of Technology, Cambridge, MA 02139, USA\\
  $^6$ CRESST, University of Maryland, Baltimore County, 1000 Hilltop
  Circle, Baltimore, MD\ 21250, USA\\ $^7$ NASA Goddard Space Flight
  Center, Astrophysics Science Division, Code 661, Greenbelt, MD\
  20771,
  USA\\$^8$ Kepler Center for Astro and Particle Physics,  Sand 1, 72076 T\"ubingen, Germany\\
  $^9$\textsl{INTEGRAL} Science Data Centre, 16 ch.\ d'\'Ecogia, 1290
  Versoix, Switzerland\\ $^{10}$ IASF--INAF, via Ugo la Malfa 153, 90136 Palermo, Italy\\
  E-mail:\\ \email{ingo.kreykenbohm@sternwarte.uni-erlangen.de}
}

\FullConference{7th INTEGRAL Workshop -- An INTEGRAL View of Compact
  Objects\\September 8--11 2008\\Copenhagen, Denmark}

\ShortTitle{Highly structured wind in Vela X-1}

\abstract{We present an in-depth analysis of the spectral and temporal
  behavior of a long almost uninterrupted INTEGRAL observation of Vela
  X-1 in Nov/Dec 2003. In addition to an already high activity level,
  Vela X-1 exhibited several very intense flares with a maximum
  intensity of more than 5 Crab in the 20--40 keV band. Furthermore
  Vela X-1 exhibited several off states where the source became
  undetectable with ISGRI.  We interpret flares and off states as
  being due to the strongly structured wind of the optical companion:
  when Vela X-1 encounters a cavity in the wind with strongly reduced
  density, the flux drops, thus potentially triggering the onset
  of the propeller effect which inhibits further accretion, thus
  giving rise to the off states. The required drop in density to
  trigger the propeller effect in Vela X-1 is of the same order as
  predicted by theoretical papers for the densities in the OB star
  winds. The same structured wind can give rise to the giant flares
  when Vela X-1 encounters a dense blob in the wind.  Further temporal
  analysis reveals that a short lived QPO with a period of ~6800 sec
  is present. The part of the light curve during which the QPO is
  present is very close to the off states and just following a high
  intensity state, thus showing that all these phenomena are
  related. }

\begin{document}

\section{Introduction}
\label{sect:intro}
Vela X-1 is a high mass X-ray binary (HMXB) consisting of the super
giant HD\,77581 and a massive (1.9\,\Msun \citep{quaintrell03a})
neutron star in a 8.964\,day orbit \citep{kerkwijk95a}. The optical
companion has a mass of $\sim$23\,\Msun and a radius of
$\sim$30\,\Rsun \citep{kerkwijk95a}.  The neutron star is deeply
embedded in the dense stellar wind of the donor \hd ($\dot M_\star = 4
\times 10^{-6}$\,\Msun$\text{yr}^{-1}$) \citep{nagase86a}. X-ray lines
indicate that this wind is inhomogeneous with many dense clumps
\citep{oskinova08a} embedded in a far thinner, highly ionized
component \citep{sako99a}.

The neutron star has a long spin period of $\sim$283\,s
\citep{rappaport75a}.  The evolution of the spin period is best
described by a random walk as expected for a wind-accreting system
\citep{ziolkowski85a}. Although the source exhibits strong
pulse-to-pulse variations, a pulse-profile folded over several pulse
periods shows remarkable stability \citep{staubert80a}, even over
decades \citep{raubenheimer90a}. At energies below 5\,keV, the
pulse-profile consists of a complex five-peaked structure, which
transforms at energies above 20\,keV into a simple double-peaked
pulse-profile \citep{staubert80a} where the two peaks are thought to
be due to the two accreting magnetic poles of the neutron star.

With an X-ray luminosity of \ca$4 \times 10^{36}\,\text{erg\,
  s}^{-1}$, \vela is a typical high mass X-ray binary.  Previous
observations have shown that the source is strongly variable with
reductions to less than 10\% of its normal value
\citep{kreykenbohm08a,kreykenbohm99a,kretschmar99a,inoue84a}, while
periods of increased activity have also been observed during which the
flux increases within an hour to a multiple of the previous value,
reaching peak flux levels close to 1\,Crab
\citep{kreykenbohm99a,haberl90a,kendziorra89a}. In this respect, \vela
is similar to sources such as 4U\,1700$-$377 and 4U\,1907$+$09, for
which low luminosity states and flares have also been observed, as is
rather typical for wind-accreting systems
\citep{fritz06a,vandermeer05a,zand97a,haberl89a}.  Although \vela is a
well studied object, only observations by \integral revealed that the
flares in \vela can be brighter than previously anticipated
\citep{kreykenbohm08a,staubert04a,krivonos03a}.

\section{Data}
\label{sect:data}

\begin{figure}
\centerline{\includegraphics[width=0.99\textwidth]{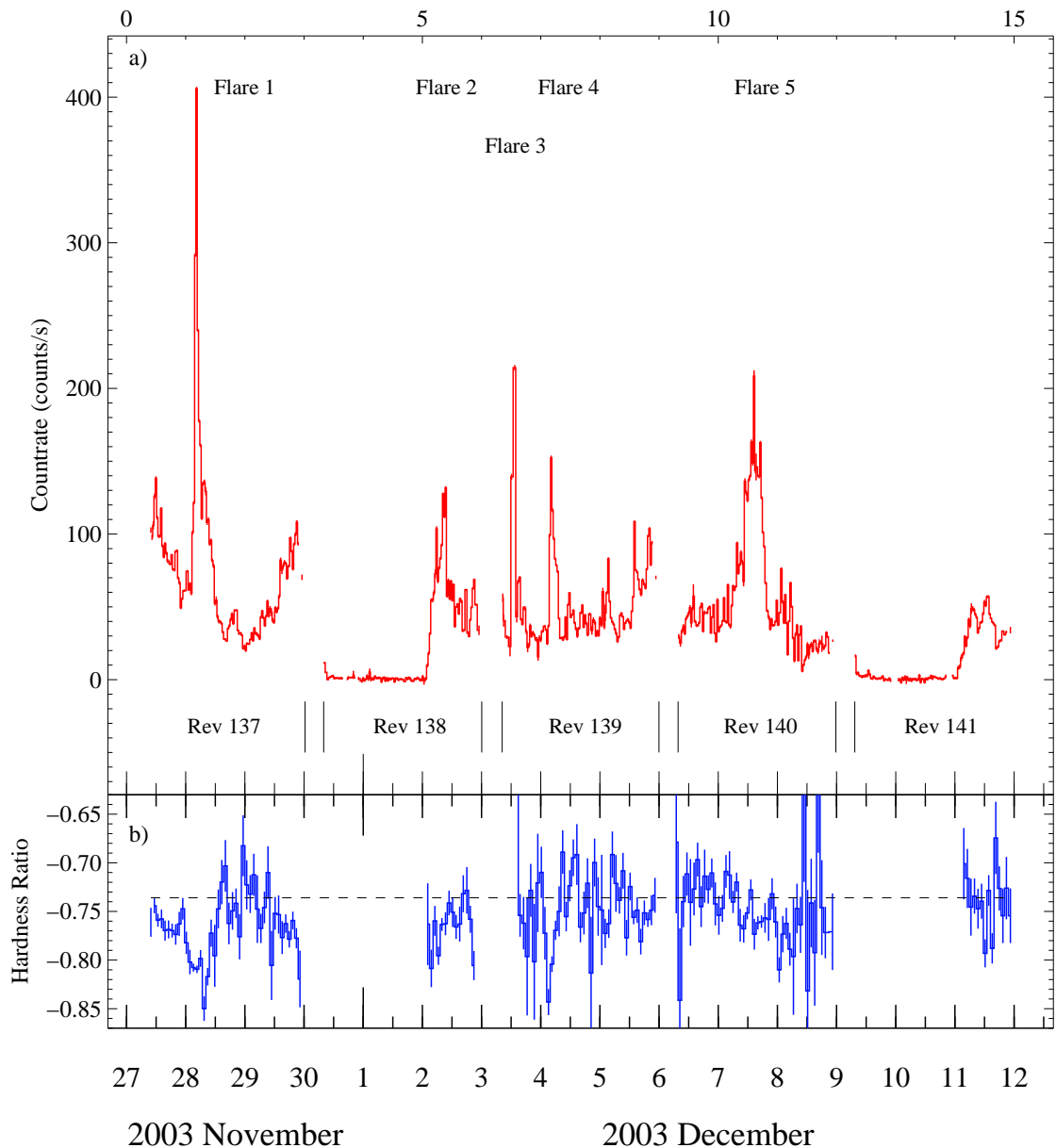}}
\vfill
\caption{\textbf{a}) ISGRI ScW by ScW light curve in the 20--40\,keV energy band
  and \textbf{b}) hardness ratio. The labels indicate the revolution
  number (from \citep{kreykenbohm08a}).}
\label{fig:lc}
\end{figure}

\integral observed the Vela region continuously for five consecutive
\integral revolutions from revolution~137 (JD\,2452970.86) until
revolution~141 (JD\,2452985.44) resulting in approximately 1\,Msec of
data (see Fig.~\ref{fig:lc}).

We used \emph{all} available science windows (ScWs) to be able to
derive a contiguous light curve with as few interruptions as
possible. Since \vela is a bright source, the OSA can detect the
source and determine its flux level accurately even when the source is
at high off-axis angles.

Apart from \vela, also 4U\,0836$-$429, H\,0918$-$5459, the
Vela~Pulsar, and two sources first reported by \integral
\citep{denhartog04a,sazonov05a} are detected. \vela is widely
separated from the other sources such that contamination of the
spectrum of \vela is of no concern. Data from JEM-X and SPI have not
been used in this analysis due to the far smaller field-of-view of
JEM-X and since \vela is off-center in the observed field (\vela was
only within the fully coded field-of-view of JEM-X for less than ten
out of the $\sim$550 individual pointings).

\section{Data analysis}
\label{sect:analysis}
\vela was found in a highly variable state during the
observation. While \vela is known to be a variable source
\citep{kreykenbohm99a,haberl94a}, the behavior found in this
observation \citep{kreykenbohm08a,staubert04a} is indeed extreme.

Most importantly, on 2003 November 28 (JD\,2452971.67), \integral
observed an extremely bright flare (flare~1; see
Fig.~\ref{fig:lc}). During the flare, the 20--40\,keV count rate
increased from a ScW averaged pre-flare value of \ca55\,\cps
($\sim$300\,mCrab, or $1.6\times 10^{-9}\,\text{erg}\,\text{cm}^{-2}\,\text{s}^{-1}$)
by a factor of more than seven to 405\,\cps (2.3\,Crab) within only
90\,minutes.

After the peak, the flare decayed quickly to an intensity level of
$<$1\,Crab and within \ca11\,h to a ScW averaged count rate of
\ca35\,\cps (200\,mCrab), somewhat lower than before the onset of the
flare (see Fig.~\ref{fig:lc}).

\begin{figure}
\centerline{\includegraphics[width=0.75\columnwidth]{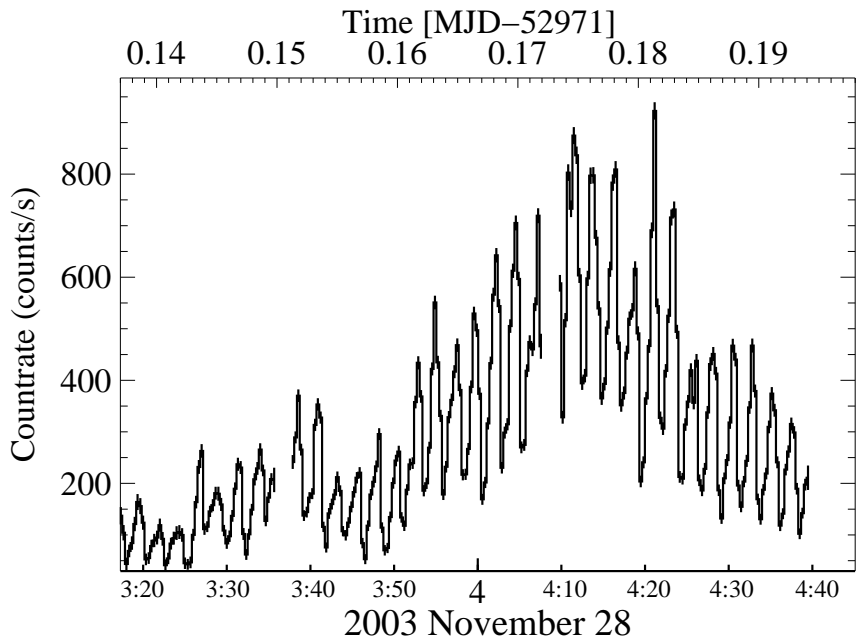}}
\caption{Close up of the light curve of giant flare~1 with a time
  resolution of 20\,s. The peak is reached at MJD\,52971.18 with
  923\,\cps (corresponding to 5.2\,Crab) in the 20 to 40\,keV band
  (from \citep{kreykenbohm08a}).
}
\label{fig:flare_lc}
\end{figure}

\begin{table}
  \caption{Overview of the observed flares. See Fig.~1 for the
    numbering of the flares. The time is the onset of the flare. To
    obtain the peak fluxes $F_\text{peak}$, a light curve with a time resolution of
    20\,s was used. $T_\text{rise}$ is the time from the onset of
      the flare to the peak, while $T_\text{total}$ is the duration of the flare.}
\label{tab:flares}
\begin{tabular}{c@{ }c@{ }r@{ }c@{ }cp{0.5\columnwidth}}
  \hline
  \hline
  Flare & Time & Duration & $F_\text{peak}$ &
  $\frac{T_\text{rise}}{T_\text{total}}$ & Remarks\rule{0pt}{1.1em} \\
  & [MJD] & \multicolumn{1}{c}{[s]} & \multicolumn{1}{c}{[Crab]} & & \\
  \hline
  1 & 52971.15 & 11\,200 & 5.2 & 0.15 & giant flare, spectral softening\\
  2 & 52975.34 &  5\,200 & 2.6 & 0.83 & no spectral change\\
  3 & 52976.50 &  1\,800 & 5.3 & 0.28 & giant flare, very short\\
  4 & 52977.15 & 12\,900 & 1.9 & 0.13 & spectral softening \\
  5 & 52980.31 & 31\,400 & 3.9 & 0.63 & high intensity state, no spectral change\\
  \hline
\end{tabular}
\end{table}

In the following 13 days, three more flares (flares~2 to~4, see
Table~\ref{tab:flares} were observed. All three flares were shorter
and less intense than flare~1 on a science window averaged basis, but
still reached ScW averaged intensities close to 1\,Crab.

On 2003 December 7 (JD\,245981.10), another intense flare was observed
(designated Flare~5, see Fig.~\ref{fig:lc}). Unlike flare~1, during
which the brightness of the source increased rapidly, it took
$\sim$8\,h for flare~5 to reach its ScW averaged maximum 20--40\,keV
flux of $\sim$1.2\,Crab. The decay lasted $\sim$5\,h until \vela
reached its pre-flare count rate of $\sim$35\,\cps (200\,mCrab in
20--40\,keV). Although quite bright, flare~5 is therefore
significantly less intense than giant flare~1, and also far longer,
i.e. it is a high intensity state.


The analysis of a light curve with a 20\,s time resolution showed that
the source reached a peak count rate of \ca920\,\cps (5.2\,Crab) in
flare~1 and \ca930\,\cps (5.3\,Crab) in flare~3. Flare~3 on December
3 was therefore also a giant flare. However, flare~3 was significantly
shorter: the entire flare lasted less than 2000\,s, but it was as
bright as flare~1 (see Table~\ref{tab:flares}).

\begin{figure}
\centerline{\includegraphics[width=0.75\columnwidth]{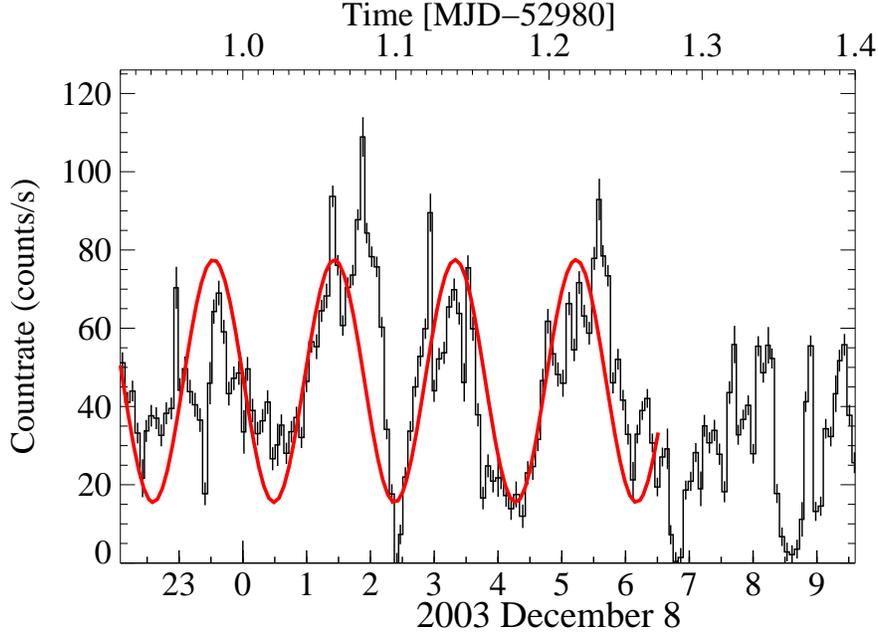}}
\caption{Closeup of the light curve (with a time resolution of 283\,s
  to remove the pulsations) where the temporary QPO is present.  Note
  that during the trough between 2\,h and 3\,h, and especially
  following the quasi-periodic modulation, the count rate decreased
  several times to zero for a short time (from
  \citep{kreykenbohm08a}).}
\label{fig:qpo}
\end{figure}

Extending the analysis to the non-flaring parts of the light curve, we
detected a quasi-periodic oscillation (QPO), similar to other
accreting X-ray pulsars. The short-lived QPO with a period of
\ca6820\,sec appears to be quite regular and inconsistent with pure
stochastic behavior (see Fig.~\ref{fig:qpo}).  Subsequent period
searches on the corresponding data subset clearly detect the period.
We note that the quasi-periodic modulation shown in Fig.~\ref{fig:qpo}
is far stronger and inconsistent with the NOMEX effect, which can
cause intensity variations from pointing to pointing, but not within a
given pointing.

\begin{figure}
\centerline{\includegraphics[width=0.75\columnwidth]{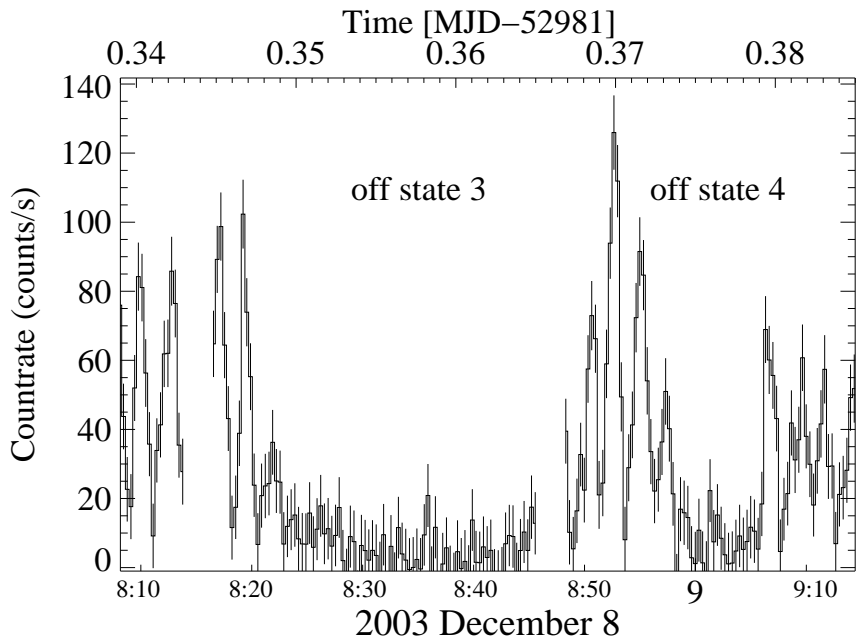}}
\caption{Close-up with a time resolution of 20\,sec on off states~3
  and~4 of \vela during which the source becomes undetectable by ISGRI
  and then turns on again within one hour (from
  \citep{kreykenbohm08a}). }
\label{fig:offstate}
\end{figure}

Furthermore, we observed several off states, during which no
significant residual flux was detectable by ISGRI
(Fig.~\ref{fig:offstate}).  The onset of these off states occured very
suddenly.  The luminosity of the source simply drops below the
detection limit of ISGRI. At the end of the off states, \vela switches
instantly on again and immediately resumes its normal intensity level.
All off states occured within 12\,h from MJD\,52981.0 and
MJD\,52981.5.

\section{Spectral evolution during the flares}
\label{Sect:hardness}
Although the source was extremely bright during the flares, meaningful
spectral fits could not be obtained, as the exposure time was to
short. We therefore analyze the hardness ratio  (see
Fig.~\ref{fig:lc}b), which is defined as
\begin{equation}\label{eq:hardness}
\text{HR} = \frac{H-S}{H+S}
\end{equation}
where $H$ is the count rate in the hard band (40--60\,keV) and $S$ the
count rate in the soft band (20--30\,keV). While the hardness ratio
remained constant throughout most of the observation at $\sim-0.735$,
the hardness ratio significantly changed with the onset of flare~1: it
dropped to $-0.82$ and during the flare to $-0.85$ (see
Fig.~\ref{fig:lc}b). The same for flare~4: the hardness ratio dropped
from  $-0.72$ to $-0.84$, the same level as in giant flare~1,
although flare~4 was far shorter and reached only a third of the peak
flux of flare~1. During flares~2 and~5, however, the hardness ratio
did not change.

\begin{figure}
\centerline{\includegraphics[width=0.75\columnwidth]{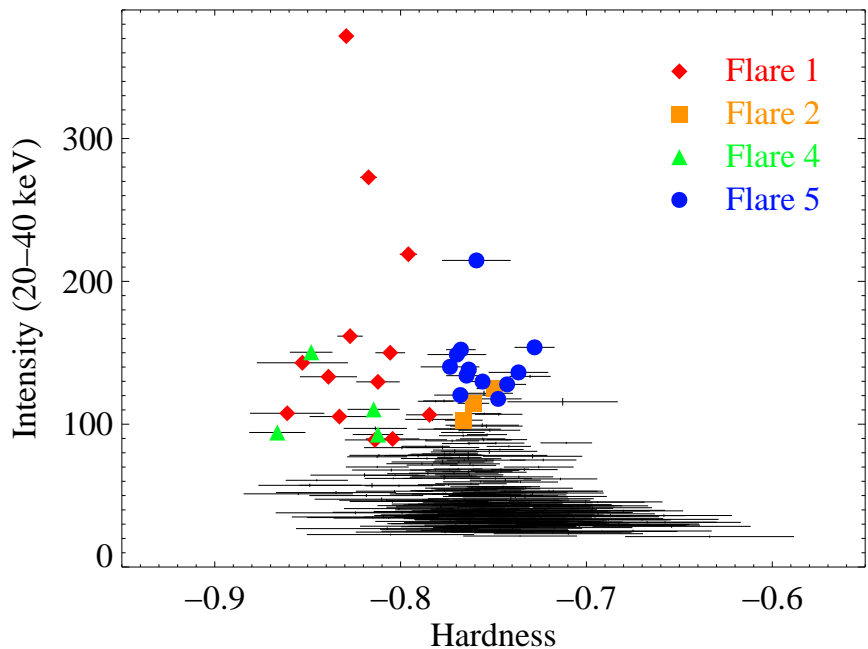}}
\caption{Hardness Intensity Diagram of \vela; data from the eclipses
  have been excluded.  The data points from the flares are indicated
  by individual symbols. The datapoints from the eclipses have been
  removed (from \citep{kreykenbohm08a}).  }
\label{fig:hid}
\end{figure}

We then used ``Hardness intensity diagrams'' (HIDs) to study the
spectral evolution of the source (see Fig.~\ref{fig:hid}). Most of the
data points are centered around the average values of intensity and
hardness ratio. The only exception are the data points from the
flares, which are above the general cluster of data points. Due to the
spectral softening, the data points of flare~1 and flare~4 are
shifted. The softening, however, did not evolve during the flares, but
the flares are softer than the average spectrum from the beginning
until the end.

\section{Discussion}
\label{sect:discussion}

\subsection{The flares}

\vela has always been known to be highly variable with time and to
show intensity variations of up to a multiple or a fraction of the
original intensity on all time scales.    Although \vela
has exhibited extensive flaring activity in the past however, giant
flares (as flares~1 and~3) had not been seen before.

The analysis of the hardness ratio shows that there seem to be two
types of flares: the first type (flares~1 and~4) shows dramatic
increases in the count rate, the onset of the flare is very sudden,
and the spectrum softens during the flare. The second type is similar
to a high intensity state: these flares are longer and the spectrum
does not change.

The mechanism behind the flaring activity, however, is not fully
understood. It has been shown that a temporary disk may form in \vela
\citep{taam89a}. The disk collapses and the material is accreted onto
the neutron star giving rise to a short flare. These predicted flares
would last from 15 to 60 minutes, similar to the short flaring
activity of \vela. Furthermore, wind accretion is a highly instable
process by itself: the accretion wake trailing the neutron star
contains filaments which also produce flares when being accreted
\citep{blondin90a}. The shock trailing the neutron star oscillates
creating the ``flip-flop instability'' which then produces inflows
that repeatedly change their direction
\citep{matsuda91a,benensohn97a}. The timescale of 45\,min matches some
of the observed behavior very well, but fails to explain the long
flares.

As the local density in a shocked wind varies by a factor of 100
\citep{kaper93a}, which can explain the flaring X-ray luminosity
\citep{oskinova08a},  dense clumps trapped in an otherwise thin and
more homogeneous wind might be responsible for long flares
\citep{leyder07a}, when the clumps are being accreted. Such a clump
can feed the neutron star with a significantly higher $\dot M$ than
usual over several hours.  In summary, we conclude that the observed
long flares are due to a strongly structured OB star wind, while when
\vela is less active, the OB star wind is less structured.

\subsection{The off states}
\label{disc:offstates}

In a similar way to the flaring activity, the off states
\citep{inoue84a,kreykenbohm99a,kreykenbohm08a} where the source is
below the detection limit are remarkable. After the off state observed
by \textsl{RXTE} in 1996 \citep{kreykenbohm99a}, the source resumed
its normal, pulsating behavior without any transition phase. The
offstates reported here also occured without a transition phase
\citep{kreykenbohm08a}.

The reasons for these off states and the sudden reappearance of
pulsations are not understood. So far, several ideas to explain these
phenomena have been proposed, however, none can fully explain the
observed off states, since all of these ideas require a significantly
longer timescale, e.g. clumps in the stellar wind \citep{feldmeier97a}
would have to have an unrealistic high optical depth to completely
block hard X-rays and can not pass the line of sight in a few seconds
to explain the sudden turn on/off behavior of the source. Therefore
other mechanisms must be considered.

The wind of OB super giants is inhomogeneous and clumpy
\citep{walter07a,blondin90a}. The density of the stellar wind can vary
by several orders of magnitude \citep{runacres05a}. Not only clumps,
but also holes, i.e.  regions of strongly reduced density are present
in the wind: there, the density is lower than the average density of
the wind by a factor of $10^3$ \citep{runacres05a}. If the neutron
star enters these holes, $\dot M$ would then also decrease by a factor
of $\sim10^3$ and the X-ray luminosity would be reduced
accordingly. Furthermore, the density fluctuations predicted by these
models occur suddenly \citep[see Fig.~1 in][]{runacres05a} similar to
the onset of the off states (see Fig.~\ref{fig:lc}). If $\dot M$ drops
due to these density variations in the wind, the Alfv\'en radius will
increase due to the reduced ram pressure of the infalling gas. Once it
is larger than the co-rotation radius, accretion onto the neutron star
is inhibited, i.e. the X-ray source basically switches off. This
scenario is commonly known as the propeller effect
\citep{illarionov75a}.  Since the propeller effect depends on the
amount of infalling material, the Alfv\'en radius is not constant.
This effect was observed in GX\,1$+$4 \citep{cui97b}: in very low
luminosity states, no pulsations were observable, while the source was
strongly pulsating in high luminosity states.

Since the strength of the magnetic field of \vela is known from the
observation of the cyclotron lines \citep{kreykenbohm02b}, the
critical flux limit for \vela for the onset of the propeller effect
can be obtainted (after \citep{cui97b}): $F_\text{X,Propeller,\vela}
\approx 1.1 \times 10^{-12}\,\text{erg cm}^{-2}\text{s}^{-1}$.
Compared with the typical bolometric flux of several times $10^{-9}$\,erg
cm$^{-2}$s$^{-1}$, this critical flux is lower by about three orders of
magnitude. This flux limit matches very closely the predicted density
variations in the stellar wind of $10^{3-5}$ \citep{walter07a}.

We therefore conclude that off states could be caused by a sudden drop
in $\dot M$ that allows \vela to enter the propeller regime. Intensity
dips, however, are longer, show a smooth transition, and exhibit
photoelectric absorption of more than $10^{24}$\,cm$^{-2}$. These dips
are readily explained by a dense blob in the wind passing through the
line of sight.

\subsection{Connection with SFXTs}

The similarity between the flares and off states in \vela and the
behavior of Supergiant Fast X-ray Transients (SFXTs) \citep{sguera05a}
is striking.  SFXTs are high mass X-ray binaries that show very brief
outbursts on timescales of hours or even only tens of minutes, and
then remain undetectable at higher energies for months between
outbursts \citep{negueruela08a}.  SFXTs should be rather bright
persistent objects \citep{grebenev07a} since the neutron star is
deeply embedded in the dense stellar wind of the optical companion,
however, the accretion is inhibited by the propeller
effect. It has therefore been proposed that SFXTs harbor a
magnetar \citep{bozzo08a} and that the extremely strong magnetic field ($B >
10^{15}$\,G) effectively inhibts accretion unless the ram pressure of
the infalling gas is high enough such that accretion becomes possible
for a short time.  The giant flares and off states of \vela are
therefore similar to these outbursts: in both cases the accretion of a
dense blob of material causes the outburst or flare, while reduced
material infall causes the source to switch off.  In summary, \vela
and SFXTs are rather similar objects, however, SFXTs are usually in
the off state, while \vela is usually in a normal accretion mode.


\bibliographystyle{PoS}
\bibliography{mnemonic,aa_abbrv,ikabbrv,velax1,div_xpuls,xpuls,cyclotron,books,roentgen,satelliten,foreign,misc}

\end{document}